\documentclass[aps,prl,reprint,groupedaddress]{revtex4-1}
\pdfoutput=1 
\usepackage{graphicx}
\usepackage{color}

\newcommand{\beq}{\begin{equation}}
\newcommand{\eeq}{\end{equation}}
\newcommand{\bea}{\begin{eqnarray}}
\newcommand{\eea}{\end{eqnarray}}

\begin{document}
\title{Freeze-out temperature from net-Kaon fluctuations at RHIC}

\author{R. Bellwied$^a$, J. Noronha-Hostler$^b$, P. Parotto$^a$, I. Portillo Vazquez$^a$, C. Ratti$^a$, J. M. Stafford$^a$}
\affiliation{\small{\it $^a$ Department of Physics, University of Houston, Houston, TX, USA  77204}}
\affiliation{\small{\it $^b$ Department of Physics and Astronomy, Rutgers University, Piscataway, NJ USA 08854}}

\date{\today}
\
\begin{abstract}
We compare the mean-over-variance ratio of the net-kaon distribution calculated within a state-of-the-art hadron resonance gas model to the latest experimental data from the Beam Energy Scan at RHIC by the STAR collaboration.  Our analysis indicates that it is not possible to reproduce the experimental results using the freeze-out parameters from the existing combined fit of net-proton and net-electric charge mean-over-variance. The strange mesons need about 10-15 MeV higher temperatures than the light hadrons at the highest collision energies. In view of the future $\Lambda$ fluctuation measurements, we predict the $\Lambda$ variance-over-mean and skewness-times-variance at the light and strange chemical freeze-out parameters. We observe that the $\Lambda$ fluctuations are sensitive to the difference in the freeze-out temperatures established in this analysis. Our results have implications for other phenomenological models in the field of relativistic heavy ion collisions. 
\end{abstract}
\pacs{}

\maketitle

\section{Introduction}

Relativistic heavy ion collisions have successfully recreated the Quark Gluon Plasma (QGP) in the laboratory at the Relativistic Heavy Ion Collider (RHIC) and the Large Hadron Collider (LHC).  At low baryon density, the transition from a deconfined state where quark and gluons are the main degrees of freedom into the hadron gas phase where quarks and gluons are confined within hadrons is a smooth cross-over that happens across roughly a 20 MeV range in temperature \cite{Aoki:2006we,Borsanyi:2010bp,Bazavov:2011nk}.  At large enough baryon densities one expects a critical point to eventually be reached; searches for such a critical point are the focus of the Beam Energy Scan at RHIC.

The evolution of a heavy ion collision is characterized by several steps, which can be described by different theoretical approaches. The chemical freeze-out is the moment at which inelastic collisions between particles cease. The chemical composition of the system is fixed at this point. Therefore, measured particle multiplicities and fluctuations carry information about this particular moment in the evolution of the system. Using thermal fits, where one assumes a hadron resonance gas in equilibrium, one can calculate the particle yields of hadrons in a $\{T,\mu_B\}$ plane and then compare to experimental data to extract the corresponding  $\{T_{f},\mu_{Bf}\}$ at the chemical freezeout. The Hadron Resonance Gas (HRG) model has been very successful in fitting particle yields and ratios over nine orders of magnitude \cite{Cleymans:1998yb,Andronic:2008gu,Manninen:2008mg,Abelev:2009bw,Aggarwal:2010pj}.
However, at small baryon densities there still remains a tension between the yields of light particles versus strange particles \cite{Floris:2014pta}. Analyzing the most recent experimental data at the LHC, it appears that light hadrons prefer a smaller chemical freeze-out temperature compared to strange hadrons (or put in other words, the strange baryons are underpredicted if the chemical freeze-out temperature is set by the light hadrons). A similar effect was observed at RHIC \cite{Adamczyk:2017iwn}.

There have been a number of suggestions to explain the tension betweeen light and strange hadrons.  For instance, from lattice QCD there is an indication that strange particles might hadronize at a higher temperature \cite{Bellwied:2013cta}, naturally leading to a higher chemical freeze-out temperature as well. It is now possible to test this idea more thoroughly by taking the moments of the net-$K$ distribution (due to the fluctuations of the number of kaons to anti-kaons on an event-by-event basis) and comparing them directly to Lattice QCD \cite{Noronha-Hostler:2016rpd}, although results for the full Beam Energy Scan from Lattice QCD are not yet available.  Another suggestion has been that missing resonances could explain the difference in temperatures \cite{Bazavov:2014xya,Noronha-Hostler:2014aia,Noronha-Hostler:2014usa}. While recent Lattice QCD calculations \cite{Bazavov:2014xya,Alba:2017mqu} indicate that several resonances are indeed missing, their full decay channels need to be implemented in order to determine their influence on the freeze-out temperature. Furthermore, previous studies found that the inclusion of additional resonances did not have a significant influence on the thermal fits \cite{NoronhaHostler:2009tz}. Another idea would be that, due to the large annihilation cross sections, (anti-)proton freeze-out is expected to occur at lower temperatures \cite{Becattini:2016xct,Rapp:2000gy,Rapp:2001bb,Rapp:2002fc,Steinheimer:2012rd,Becattini:2012xb} and thermal equilibrium will not be reached within the hadronic fireball.

A clear path forward to tackle this tension between light and strange hadrons is to study the moments of the light hadron distributions (namely net-protons and net-charge) as well as the net-kaon distribution. For certain particles (including kaons) fluctuations turn out to be more sensitive to the freeze-out parameters, compared to the corresponding particle yields \cite{Alba:2015iva}. The STAR collaboration recently published experimental measurements for the energy dependence of the fluctuations of net-protons \cite{Adamczyk:2013dal}, net-charge \cite{Adamczyk:2014fia}, and net-kaons \cite{Adamczyk:2017wsl}. Experimentally, one can only measure charged particles, so that $K^0$'s, $\pi^0$'s, and neutrons are not included in these measurements. A previous study in the HRG model with all experimental effects, such as acceptance cuts in $p_T$ and rapidity and isospin randomization \cite{Alba:2014eba}, found that the net-proton and net-charge fluctuations indicate a lower chemical freeze-out temperature than the one quoted in the thermal fits. 
%

If chemical freeze-out is reached within the hadron gas phase, then only hadronic degrees of freedom should be considered, which makes the HRG model the ideal tool to study this point in the evolution of the system.  An advantage in using the HRG model is that acceptance cuts and resonance decays can be taken into account \cite{Noronha-Hostler:2016rpd}, which is not possible when directly comparing to Lattice QCD results.  While these effects appear to be small at high collision energies \cite{Noronha-Hostler:2016rpd}, at large baryon chemical potentials and for higher order susceptibilities they do play a role. In this paper we use the HRG model to extract the kaon chemical freeze-out parameters by comparing the model predictions for net-kaon fluctuations to the recent STAR data from the Beam energy scan \cite{Adamczyk:2017wsl}. We find that, even when using the most up-to-date particle data list as an input for the model, the kaons need larger freeze-out temperatures, compared to the light hadrons. We also predict the values of the $\Lambda$ fluctuations, calculated in the HRG model at the freeze-out parameters of the kaons and of the light hadrons. The results show a clear separation, which can hopefully be resolved by the forthcoming experimental results. 

\section{Methodology}

\begin{figure*}[t]
\centering
\includegraphics[width=\linewidth]{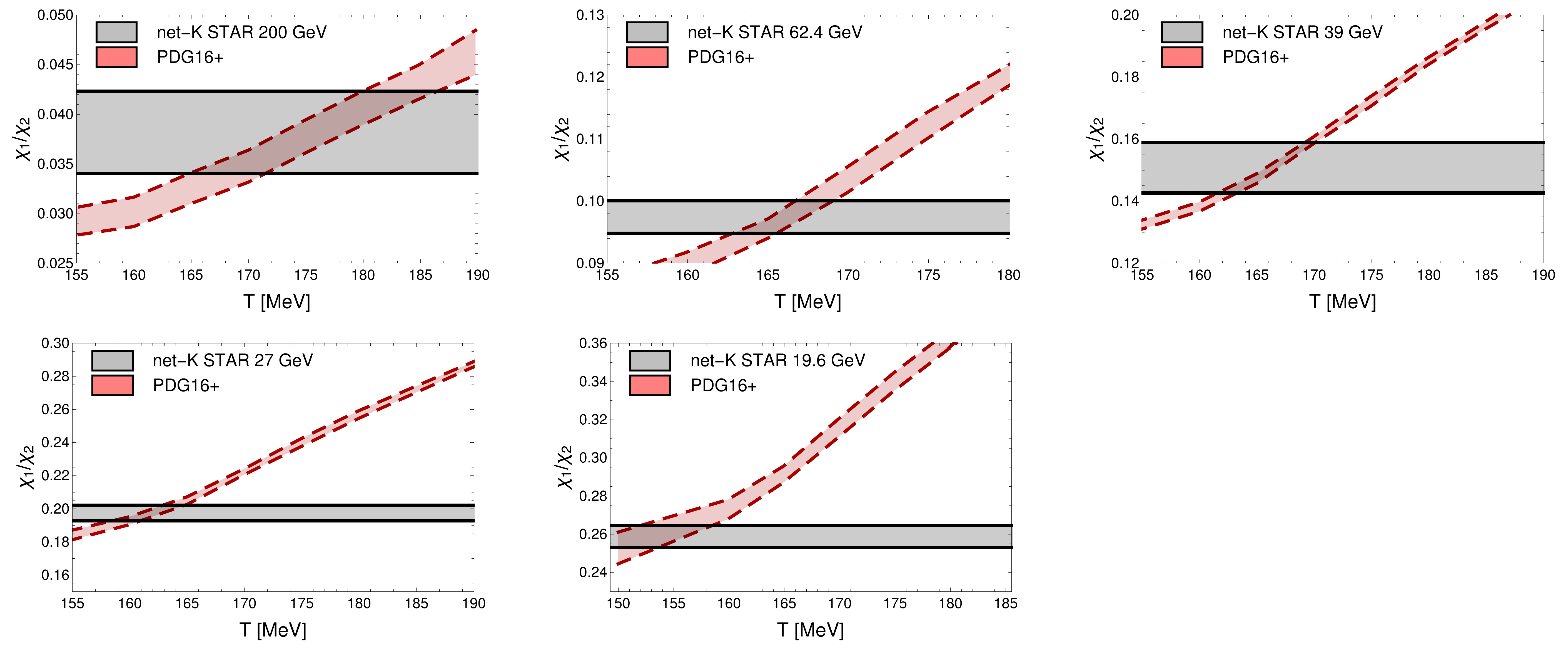} 
\caption{Results for $\chi_1^K/\chi_2^K$ calculated along the Lattice QCD  isentropic trajectories  (pink band) compared to  $(M/\sigma^2)_K$ data from \cite{Adamczyk:2017wsl} (gray band) across the Beam Energy Scan at STAR.} \label{fig:isen}
\end{figure*}

The HRG model  assumes that a gas of interacting hadrons in its ground state can be well approximated by a non-interacting gas of ground-state hadrons and their heavier resonances.  The pressure from this model is defined as
\begin{widetext}
\begin{equation}
p(T,\mu_B,\mu_Q,\mu_S) =  \sum_i (-1)^{B_i+1} \frac{d_iT}{(2\pi)^3} \int {\rm d}^3\vec{p} \,\,\ln\Big[1+(-1)^{B_i+1} \exp({-(\sqrt{\vec{p}^{\,2}+m_i^2}-B_i\mu_B-S_i\mu_S-Q_i\mu_Q)/T})\Big] \,,
\label{equ:pressureHRG}
\end{equation}
\end{widetext}
namely it is the sum over all known baryons and mesons of the pressure of a baryon/meson gas. The conserved charges are baryon number $B$, strangeness $S$, and electric charge $Q$.  The main input to the model is the list of hadrons that have an individual degeneracy $d_i$, mass $m_i$, and quantum numbers $B_i$, $S_i$, and $Q_i$. The chemical potentials are all linked due to strangeness neutrality and the approximate ratio of  0.4 protons to baryons in the colliding nuclei such that:
\begin{eqnarray}
\label{conditions}
\sum_{i\in S}n_i(T,\mu_B,\mu_Q,\mu_S)&=&0\nonumber\\
\sum_{i\in Q}n_i(T,\mu_B,\mu_Q,\mu_S)&=&0.4\,\sum_{i\in B}n_i(T,\mu_B,\mu_Q,\mu_S) \,.
\end{eqnarray}

In this paper we use the hadron resonance gas model with the Particle Data Group list from 2016+ including all one-star to four-star resonances, which appears to have the best global fit to the Lattice QCD partial pressures \cite{Alba:2017mqu}. The decays of these resonances are taken from the Particle Data Group directly when provided. In cases where the decays are not known, information is extrapolated from similar resonances with the same quantum numbers. For the heaviest resonances, it is assumed that, during their decay, a large portion of their rest mass goes into radiation instead of hadrons. Further details are outlined in Ref. \cite{Alba:2017hhe}. 

In order to take into account the influence of hadron resonance decays, we adapt the formula used in \cite{Alba:2014eba,Alba:2015iva}, which contains the proper acceptance cuts in rapidity and transverse momentum that were used in the experiment, to just include the decays into charged kaons:
\begin{widetext}
\begin{equation}\label{eqn:chis}
\tilde{\chi}_n^{K^{\pm}} \! \! = \! \! \sum_i^{N_{HRG}} \! \! \left(Pr_{i\rightarrow K^{\pm}}  S_{K^\pm}\right)^n \frac{d_i}{4\pi^2}\frac{\partial^n}{\partial \mu_S^n}\left\{ \int_{-0.5}^{0.5} {\rm d}y \int_{0.2}^{1.6} {\rm d}p_T\times \frac{p_T\sqrt{p_T^2+m_k^2}{\rm Cosh}[y]}{(-1)^{B_k+1}+\exp({({\rm Cosh}[y]\sqrt{p_T^{\,2}+m_k^2}-(B_i\mu_b+S_i\mu_S+Q_i\mu_Q))/T})}  \right\}.
\end{equation}
\end{widetext}
where $S_{K^\pm}$ is the strangeness of the stable particle ($\pm 1$ for $K^\pm$ in our case). Here $Pr_{i\rightarrow K^{\pm}}=Br_{i\rightarrow K^{\pm}} n_i(K^{\pm}) $ is the probility for a resonance $i$ to decay into a charged kaon where $Br_{i\rightarrow K^{\pm}}$ is the branching for the resonance $i$ to decay into $K^{\pm}$ and $n_i(K^{\pm})$ is the number of times particle $i$ appears in the channel $K^{\pm}$. Notice that $(Pr_{i\rightarrow K^\pm} S_{K^\pm})^n$ in the above equation corresponds to the sum of two terms, the first including the probability for particle $i$ of decaying into $K^+$, the second the probability of decaying into $K^-$: $(Pr_{i\rightarrow K^{\pm}}S_{K^\pm})^n = (Pr_{i\rightarrow K^{+}}S_{K^+} + Pr_{i\rightarrow K^{-}}S_{K^-})^n =  (Pr_{i\rightarrow K^{+}} - Pr_{i\rightarrow K^{-}})^n $. Here we use the same acceptance cuts as described in \cite{Adamczyk:2017wsl}.

When making comparisons to experimental data, the ratios of susceptibilities are always used to cancel out volume effects.  Then, one can calculate $\chi_1^{K^{\pm}}/\chi_2^{K^{\pm}} (T,\mu_B)$ across the entire phase diagram of temperature and baryon chemical potential. In the Beam Energy Scan different center of mass energies, that correspond to different trajectories across the QCD phase diagram (lower energies correspond to larger $\mu_B$), are systematically scanned. 

For the light hadrons, at each inidividual energy there are two unknowns: $\{T_f,\mu_{Bf}\}$ and two experimental data points to match $\{\chi_1^p/\chi_2^p,\chi_1^Q/\chi_2^Q\}$. Such an analysis was performed in Ref. \cite{Alba:2014eba}. However, for strange particles only net-kaons have been measured so it is not possible to determine both  $\{T_f,\mu_{Bf}\}$  by fitting $\chi_1^K/\chi_2^K$. We tried a simultaneous fit of $\chi_1^K/\chi_2^K$ and $\chi_3^K/\chi_2^K$, but the experimental error-bars on the latter did not allow a precise determination of the freeze-out parameters. In Fig. 3 of Ref. \cite{Gunther:2016vcp}, isentropic trajectories using Lattice QCD results for the Taylor reconstructed QCD phase diagram at finite $\mu_B$ are shown. These trajectories assume that the entropy per baryon number is conserved and illustrate the path across which the Quark Gluon Plasma evolves through the phase diagram after a heavy-ion collision in the absence of dissipation.  They are a reasonable approximation of the actual ones over a short section of the system evolution, close to the freeze-out. Thus we assume that the evolution of the system created in a heavy ion collision lies on the Lattice QCD isentropic trajectories, which yield a relationship between $T$ and $\mu_B$. 
These isentropes were determined by starting from the chemical freeze-out points for light hadrons from Ref. \cite{Alba:2014eba}, calculating $S/N_B$ at those points, and imposing that the ratio is conserved on the corresponding trajectory. In this way we take into account the possibility that kaons can freeze-out at a different moment in the evolution of the system at a given collision energy, related to the light particle freeze-out point by the conservation of $S/N_B$. This procedure allows us to determine both $\{T_f,\mu_{Bf}\}$ for Kaons.
\section{Results}
In Fig.\ \ref{fig:isen}, $\chi_1^K/\chi_2^K$ is calculated along the Lattice QCD  isentropic trajectories (pink band) and compared to the $(M/\sigma^2)_K$ (mean-over-variance) data from the STAR collaboration \cite{Adamczyk:2017wsl} (gray band).  At $\sqrt{s_{NN}}=200$ GeV, due to the large experimental uncertainty, the region of overlap between the theoretical band and the experimental data corresponds to a temperature range between $T\sim 163-190$ MeV, which is clearly above the light chemical freeze-out temperature $T^{f}=148\pm 6$ MeV.  At lower energies, the overlap region is smaller but it is still located around $T\sim 160$ MeV.  


\begin{figure}[h]
\centering
\includegraphics[width=\linewidth]{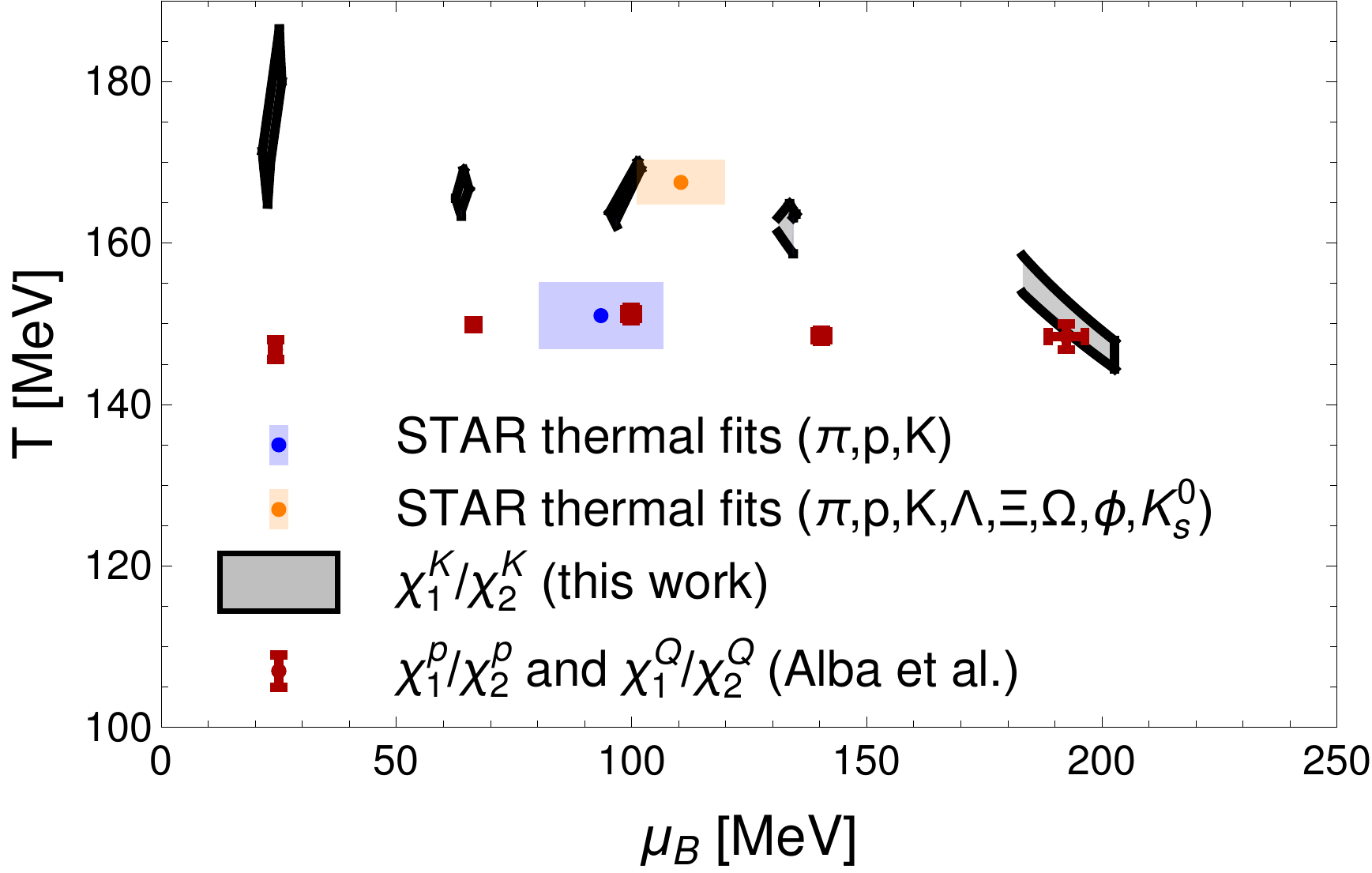} 
\caption{Freeze-out parameters across the highest five energies from the Beam energy scan. The red points were obtained from the combined fit of $\chi_1^p/\chi_2^p$ and $\chi_1^Q/\chi_2^Q$ \cite{Alba:2014eba}, while the gray bands are obtained from the fit of $\chi_1^K/\chi_2^K$ in this work. Also shown are the freeze-out parameters obtained by the STAR collaboration at $\sqrt{s}=39$ GeV \cite{Adamczyk:2017iwn} from thermal fits to all measured ground-state yields (orange point) and only to protons, pions and kaons (blue point). } \label{fig:Tmub}
\end{figure}
In Fig. \ref{fig:Tmub} we directly compare our acceptable bands for the strange $\{T^{f},\mu_B^{f}\}$ (gray bands) and the light $\{T^{f},\mu_B^{f}\}$ from Ref. \cite{Alba:2014eba} (red points). Note that the shape of the strange $\{T^{f},\mu_B^{f}\}$ regions arises from the shape of the overlap regions seen in Fig.\ \ref{fig:isen}. From the plot it is clear that, performing the same analysis as was done in Ref. \cite{Alba:2014eba} for light particles, the freeze-out parameters that we get from Kaon fluctuations are in disagreement with the light particle ones. Therefore, we conclude that the Kaon fluctuation data from the STAR collaboration cannot be reproduced within the HRG model, using the freeze-out parameters obtained from the combined analysis of $\chi_1^p/\chi_2^p$ and $\chi_1^Q/\chi_2^Q$. Kaon fluctuations seem to confirm a flavor hierarchy scenario. In the same Figure, we also show the freeze-out parameters obtained from thermal fits to particle yields by the STAR collaboration at $\sqrt{s}=39$ GeV \cite{Adamczyk:2017iwn}. The orange point has been obtained by fitting all measured ground-state hadrons, while for the blue point the fit only included protons, pions and kaons. It is clear that the inclusion of all strange particles drives the freeze-out temperature to values which are close to the ones we find from kaon fluctuations. The fit to protons, pions and kaons yields a freeze-out temperature compatible to the one obtained from the combined fit of net-proton and net-charge fluctuations in \cite{Alba:2014eba}.

\begin{figure}[h]
\centering
\includegraphics[width=\linewidth]{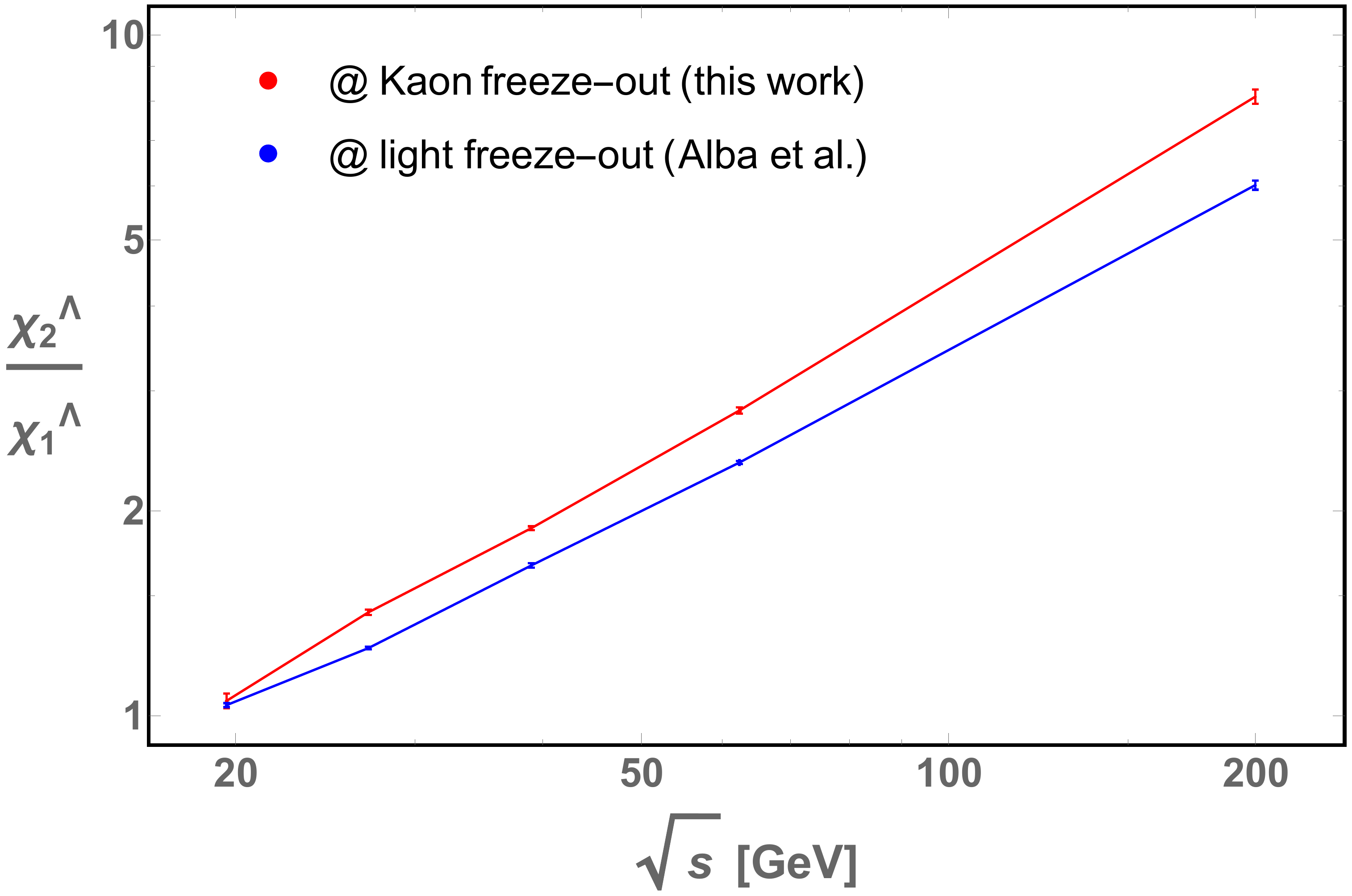} \\
\includegraphics[width=\linewidth]{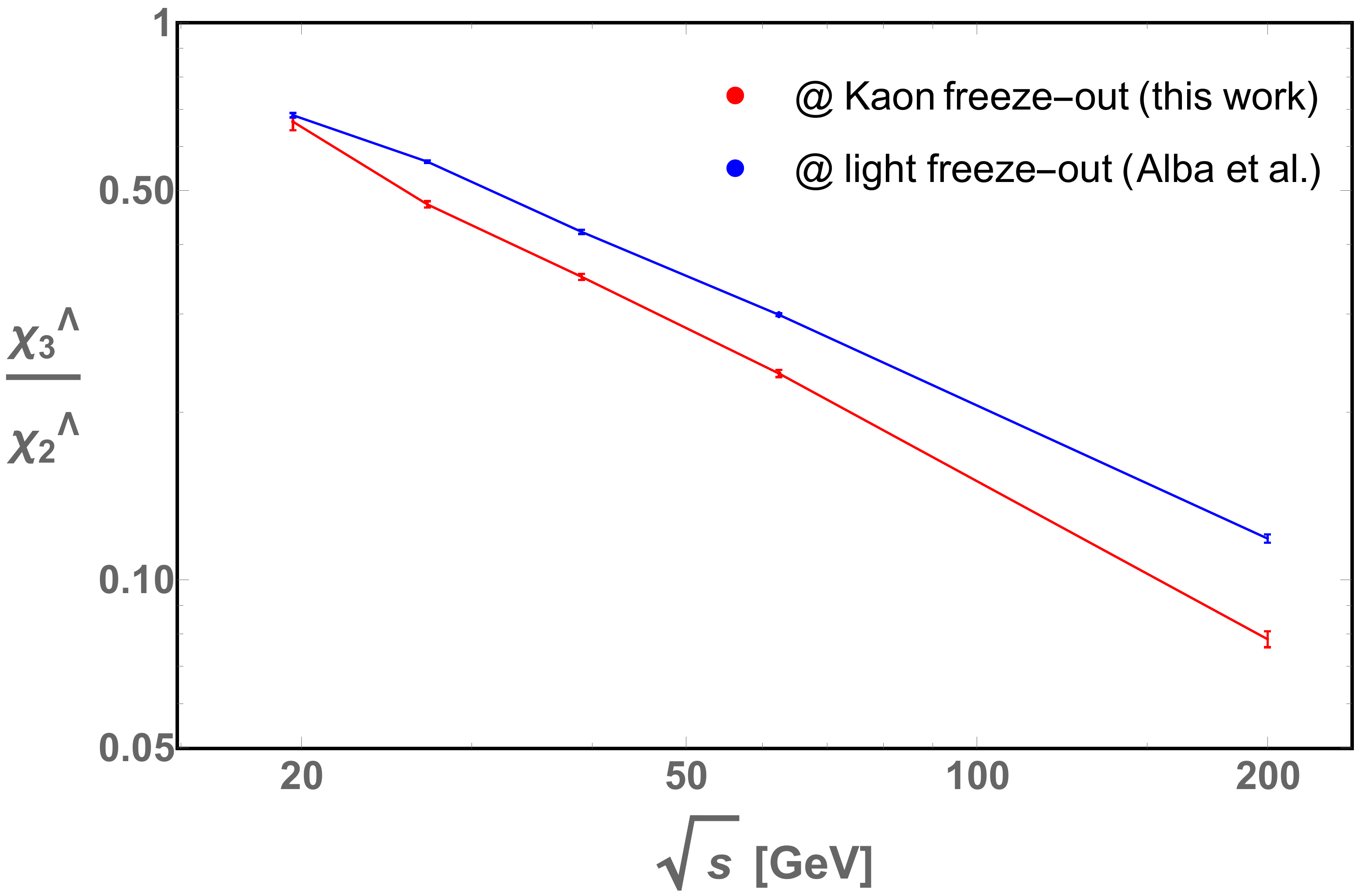}
\caption{Upper panel: $\chi_2^{\Lambda}/\chi_1^{\Lambda}$ as a function of $\sqrt{s}$. Lower panel: $\chi_3^{\Lambda}/\chi_2^{\Lambda}$ as a function of $\sqrt{s}$. In both panels, the red points are calculated at the values of $T_f$ and $\mu_{Bf}$ extracted from the fit of $\chi_1^K/\chi_2^K$, while the blue points are calculated at the values of $T_f$ and $\mu_{Bf}$ extracted from the combined fit of $\chi_1^p/\chi_2^p$ and $\chi_1^Q/\chi_2^Q$ in Ref. \cite{Alba:2014eba}. } \label{Lambda}
\end{figure}

Experimental data for $\Lambda$ fluctuations will soon become available. They could serve as a further test for the two freeze-out scenario, as they carry strangeness as well. For this reason, in Fig. \ref{Lambda} we show our predictions for $\chi_2^{\Lambda}/\chi_1^{\Lambda}$ (upper panel) and $\chi_3^{\Lambda}/\chi_2^{\Lambda}$ (lower panel) as functions of the collision energy, calculated at the values of $T_f$ and $\mu_{Bf}$ extracted from the fit of $\chi_1^K/\chi_2^K$ (red points), and from the combined fit of $\chi_1^p/\chi_2^p$ and $\chi_1^Q/\chi_2^Q$ (blue points). Both observables show a clear separation between the two scenarios, that the future experimental results will hopefully be able to resolve.

\section{Conclusions}
We performed a fit of the net-Kaon $\chi_1/\chi_2$ data from the STAR collaboration, in order to extract the freeze-out parameters for Kaons.
We observe a clear separation between the freeze-out temperatures extracted from net-kaon fluctuations and those obtained from a combined fit of net-proton and net-charge fluctuations, up to $\mu_B\sim 0-200$ GeV. Our HRG model includes the most up-to-date particle data list that works reasonable well compared to Lattice QCD results \cite{Alba:2017mqu}; their full decay channels are considered as well.  As $\mu_B$ increases, it appears that there could be a convergence of the strange vs. light temperatures; however, the acceptable band at $\sqrt{s_{NN}}=19.6$ GeV for the strange chemical equilibrium temperature is quite large, due to the current experimental and theoretical uncertainties. Thus one cannot make a clear statement at low energies. We also would like to point out that, at the highest collision energy, the overlap of the data with the isentropic trajectories is so large that it yields values of the freeze-out temperature as high as 190 MeV, which is clearly incompatible with the temperature predicted for the chiral phase transition on the lattice.

Our results have interesting implications for hydrodynamical modeling at the Beam Energy Scan. They provide the first experimental evidence (beyond tantalizing hints from thermal fits) that strange hadrons could freeze-out at around $T\sim 10-15$ MeV higher temperatures than light hadrons. Certainly, at the highest RHIC energies hydrodynamical models should attempt to freeze-out strange hadrons at higher temperatures than light hadrons.  As one goes to higher baryon chemical potentials, such an approach may no longer be needed. Finally, it is not yet clear whether strange baryons would freeze-out with the Kaons or with the light particles. For this reason, we predict the $\Lambda$ fluctuations in the two scenarios: the future experimental data will hopefully help to resolve this issue. Besides, once the $\Lambda$ fluctuation measurements become available it will be possible to perform a combined fit of $\chi_2^K/\chi_1^K$ and $\chi_2^\Lambda/\chi_1^\Lambda$, instead of relying on the isentrope constraint. This might yield a more precise determination of the strangeness freeze-out temperature.

Another interesting implication from our results is that the eventual convergence of the light and strange freeze-out temperature could be a signature of an approaching critical point.  With a cross-over phase transition, it is not surprising to find different chemical freeze-out temperatures for different conserved currents.  However, at a first order phase transition all charges are expected to freeze-out at the same temperature.

\section*{Acknowledgements}
This  material  is  based  upon  work  supported  by  the National  Science  Foundation  under  grants  no.    PHY-1654219 and OAC-1531814 and by the U.S. Department of  Energy,  Office  of  Science,  Office  of  Nuclear  Physics, within the framework of the Beam Energy Scan Theory (BEST) Topical Collaboration.  We also acknowledge the support from the Center of Advanced Computing and Data Systems at the University of Houston. The work of R. B. is supported through DOE grant DEFG02-07ER41521.

\bibliography{biblio}
\end{document}